\documentclass{article}
\usepackage{graphicx}

\title{LCG-1 Deployment and usage experience}
\author{Lev Shamardin \\ {\em shamardin@theory.sinp.msu.ru} }
\date{}
       
\begin{document}

\maketitle

\begin{abstract} 

LCG-1 is the second release of the software framework for the LHC Computing
Grid project. In our work we describe the installation process, arising
problems and their solutions, and configuration tuning details of the
complete LCG-1 site, including all LCG elements required for the
self-sufficient site. 

\end{abstract}

\vspace{1pc} 

\section{Brief introduction to LCG-1}

LHC Computing Grid (LCG) is one of the five CERN projects at the moment. The
main goal of the project is is to prepare the computing infrastructure for the 
simulation, processing and analysis of LHC data for all four of the LHC 
collaborations. This infrastructure is based on the Grid model of distributed
computing {\cite{book:grid}}. In general LCG is the {\em middleware} project, 
which means that it will hide much of the complexity of the environment from
the end user.

LCG is based on the software developed in the EU-DataGrid project
{\cite{url:edg}} and VDT {\cite{url:vdt}} which trace their roots back to the
Globus Toolkit {\cite{url:globus}}.

\section{LCG-1 Deployment scheme}

\subsection{LCG-1 Deployment overview}

The deployment model for the LCG-1 includes a central CVS repository which
stores the recommended configuration templates for the current LCG-1
release. The main deployment platform at the moment is RedHat Linux 7.3.
LCG-1 deployment team provides binary RPM packages for all LCG-1 middleware
and LHC experiments software.

The recommended installation procedure for the LCG-1 is the following. The
LCFGng configuration server is installed on the site, and the hosts running
LCG-1 services are configured automatically from the LCFGng server. 

\begin{figure}[hbt]
\begin{center}
\includegraphics[scale=0.7]{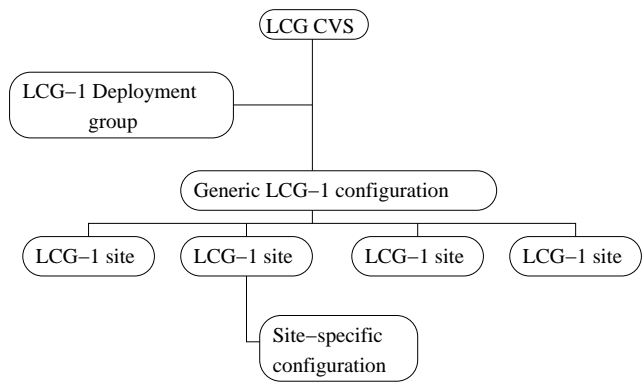}
\end{center}
\caption{LCG-1 CVS and sites configuration}
\label{fig:cvs}
\end{figure}

The configuration for the site should be based on the current LCG-1 release
version, obtained from the central CVS. For each of the sites from the 
official deployment list a module in CVS is created, and the site configuration
is then uploaded to CVS (see fig. \ref{fig:cvs}).

\subsection{LCG-1 Deployment at MOSCOW site (SINP MSU)}

As a part of the deployment test LCG-1 software was installed on the SINP MSU
PC farm (LCG-1 site name for the site is MOSCOW). The farm has 20 dual-CPU
nodes and two 1.2 TB fileservers. The fileservers have Gigabit Ethernet uplinks
and the nodes are connected with the FastEthernet network.

At the moment of the LCG-1 deployment the was an automatic software 
installation system running on our PC farm. It is based on Etherboot 
network boot package and anaconda kickstart technology. There is a 
configuration server running DHCP and TFTP services, which stores different
network boot images with preconfigured kickstart scripts for anaconda.
There is also a possibility not to use a kickstart script and to install
the node manually. In general an installation of a node from scratch requires 
a boot ROM on the node, but for the nodes without a boot ROM there is a 
possibility to boot from a special boot disk which will emulate the boot ROM.
Another possibility is to install a fake Linux kernel which also emulates a
network boot.

For the LCG-1 deployment installation system was modified in the following way.
The LCFGng server software was installed on the node running DHCP and TFTP
services, and a new network boot image for LCFGng was added. This image is based
on the standard LCFGng boot disk provided with the LCFGng software. During the
installation process the LCFGng managed node obtains the configuration
information from the server using HTTP and LCFG protocols and than installs and
configures the software packages shared on the NFS server.

Only a part of the SINP MSU PC farm was configured for LCG-1 tests. The nodes
selected for the LCG-1 were disconnected from the main farm. The following LCG-1
components were installed: Computing Element, Worker Node, Storage Element,
Resource Broker, BDII and User Interface. The MDS and MyProxy nodes will be
installed in the nearest future.

We also had some experience installing LCG-1 software on other PC farms in the
MSU. First of all, SRCC MSU parallel cluster runs its own batch system
developed by SRCC MSU and CMC MSU.  The interface between LCG middleware and
this batch system was done through the extra package to the Globus Toolkit, but
this interfaces is not yet completely tested. Other general differences between
a plain LCG-1 installation and installation on this cluster are:

\begin{itemize}
\item Manual node configuration only. The LCFGng cannot be installed on this
cluster due to some technical limitations.
\item Middleware for the Worker Nodes is installed on the shared file system.
\end{itemize}

In the nearest future we are going to install LCG middleware on the Physical
faculty and the faculty of Computing Mathematics and Cybernetics clusters in the
MSU and connect these clusters to the SINP MSU Resource Broker. However,
installation on these clusters has the similar limitations as the installation
on the SRCC MSU parallel cluster.

\section{Conclusion}

At present installation of LCG-1 middleware has both big advantages and
disadvantages.

\subsection{Difficulties}

One of the main difficulties with the LCG-1 installation for now is the lack of
available documentation. The only well-documented way to install LCG software
requires using of the LCFGng configuration server. This method is not applicable
in a number of cases.

Another issue is that the minimal standalone site must be running a big number
of nodes just for supporting the infrastructure. The minimal set of nodes
required for a standalone site is a Resource Broker, BDII \& MDS, Computing
Element with at least one Worker Node, at least one Storage Element, and the
MyProxy server if long-time jobs support is required. This gives us at least
five nodes supporting the infrastructure which cannot be used for other needs
such as running jobs for example.

\subsection{Features}

The summary of the features provided by the LCG-1 for now is the following:

\begin{itemize}
\item Convenient way for job balancing between several sites.
\item Common way of user authentication and authorization for job submission.
\item Some basic accounting over different sites.
\item Data replication.
\end{itemize}

\section{Acknowledgments}
The work was partially supported by CERN-INTAS grant 00-0440.

\end{document}